# Quantization in classical mechanics and reality of the Bohmian Ψ-field

V.D. Rusov[1,2*], D.S. Vlasenko[1], S.Cht. Mavrodiev[3,4],

[1]Department of Theoretical and Experimental Nuclear Physics,

Odessa National Polytechnic University, 65044 Odessa, Ukraine

[2]Faculty of Mathematics, Bielefeld University, P.O.X 100131, Bielefeld, Germany

[3]The Institute for Nuclear Research and Nuclear Energetics, BAS, 1874 Sofia, Bulgaria

[4]Joint Institute for Nuclear Research, Dubna, Russia

## Abstract

Based on the Chetaev theorem on stable dynamical trajectories in the presence of dissipative forces, we obtain the generalized condition for stability of Hamilton systems in the form of the Schrödinger equation. It is shown that the energy of dissipative forces, which generate the Chetaev generalized condition of stability, coincides exactly with the Bohm "quantum" potential. Within the framework of Bohmian quantum mechanics supplemented by the generalized Chetaev theorem and on the basis of the principle of least action for dissipative forces, we show that the squared amplitude of a wave function in the Schrödinger equation is equivalent semantically and syntactically to the probability density function for the number of particle trajectories, relative to which the velocity and the position of the particle are not hidden parameters. The conditions for the correctness of the Bohm-Chetaev interpretation of quantum mechanics are discussed.

**PACS number(s):** 03.65.-w, 03.65.Ta; 45.20.Jj.

___________________________________________________________

[*] Corresponding author: Rusov V.D., E-mail: siiis@te.net.ua

In this paper we consider the question, which can be formulated the following rather strict and paradoxical form: "Are the so-called quantization conditions that are imposed on the corresponding spectrum of a dynamical system possible in principle in classical mechanics, analogously to what is taking place in quantum mechanics?"

Surprisingly, the answer to this question is positive and has been given more than 70 years ago by the Russian mathematician N.G. Chetaev in his article «On stable trajectories in dynamics» [1, 2]. The leading idea of his work and of the whole scientific ideology was the most profound personal paradigm, with which begins, by the way, his principal work [2]: «Stability, which is a fundamentally general phenomenon, has to appear somehow in the main laws of nature». Here, it seems, Chetaev states for the first time the thesis of the fundamental importance of theoretically stable motions and of their relation to the motions actually taking place in mechanics. He explains it as follows: the Hamiltonian theory of holonomic mechanical systems being under the action of forces admitting the force function has well proven itself, although, as Lyapunov has shown [3], arbitrarily small perturbation forces can theoretically make such stable motions unstable. And since in actual fact holonomic mechanical systems regardless of everything often maintain stability, Chetaev puts out the paradoxical idea of the existence of special type of small perturbation forces, which stabilize the real motions of such systems [2]. Finally, Chetaev come to a conclusion that these arbitrarily small perturbation forces or, more precisely, "small dissipative forces with full dissipation, which always exist in our nature, represent a guaranteeing force barrier which makes negligible the influence of nonlinear perturbation forces" [4]. Furthermore, it has turned out that this "clear stability principle of actual motions, which has splendidly proven itself in many principal problems of celestial mechanics… unexpectedly gives us a picture of almost quantum phenomena" [5].

It is interestingly to note that a similar point of view in the different time and with the different degrees of closeness can be found in papers by Dirac [6] and 't Hooft [7]. For example, in [6] a quantization procedure appears in the framework of generalized Hamiltonian dynamics which is connected with the choice of the so-called small integrable $A$-spaces, only in which the solutions of the equations of motion, and, hence, the stable motions of a physical system are possible (see Eqs. (48) in Ref. [6]). On the other hand, according to the 't Hooft idea, the classical deterministic theory (on the Planck scale) supplemented by the mechanism of dissipation, generates the observed quantum behavior of our world on the laboratory scale. In particular, 't Hooft has shown that there is a very important class of classical deterministic systems, for which the Hamiltonian is positive due to the dissipation mechanism. This leads to "an apparent quantization of orbits which resemble a quantum structure observed in the real world [7]". It is obvious that, on the verbal level, the 't Hooft idea is a practically adequate reflection of the crux of the Chetaev idea, because the physical essence of both ideas is based on the fundamental role of dissipation in the microworld and can be characterized by a non-trivial, but unambiguous (on the Planck scale) thesis: no dissipation - no quantization!

Below we generalize the Chetaev theorem on stable trajectories in dynamics to the case where the Hamiltonian of a system is explicitly time-dependent [8]. For that let us consider a material system (where $q_1,\ldots, q_n$ and $p_1,\ldots, p_n$ are the generalized coordinates and momenta of the holonomic system) in the field of potential forces which admit a force function of $U(q_1,\ldots, q_n)$ type.

The complete intergal of the Hamilton-Jacobi differential equation corresponding to a given system has the form

$$S = f(t, q_1,\ldots, q_n; \alpha_1,\ldots, \alpha_n) + A, \tag{1}$$

where $\alpha_1,\ldots, \alpha_n$ and A are arbitrary constants, and the general solution of the mechanical problem is defined, according to the well-known Jacobi theorem, by the formulas

$$\beta_i = \frac{\partial S}{\partial \alpha_i}, \quad p_i = \frac{\partial S}{\partial q_i}, \quad i = 1,\ldots, n, \tag{2}$$

where $\beta_i$ are new constants of integration. The possible motions of the mechanical system are determined by the various values of the constants $\alpha_i$ and $\beta_i$.

We will call a motion of the material system, whose stability will be studied, the unperturbed motion. To begin with, let us study the stability of such a motion with respect to the variables $q_i$ under a perturbation of only the initial values of these variables (i.e., of values of the constants $\alpha_i$ and $\beta_i$) in the absence of disturbing forces.

If we denote the variations of the coordinates $q_j$ and momenta $p_j$ as $\xi_j = \delta q_j = q_j - q_j(t)$ and $\eta_j = \delta p_j = p_j - p_j(t)$, respectively, and the Hamilton function as $H(q_1,\ldots, q_n, p_1,\ldots, p_n)$, it is easy to obtain for the Hamilton canonical equations of motion

$$\frac{dq_j}{dt} = \frac{\partial H}{\partial p_j}, \quad \frac{dp_j}{dt} = -\frac{\partial H}{\partial q_j} \tag{3}$$

the differential equations (in the first approximation) in the Poincaré variations [9] which have the following form

$$\frac{d\xi_i}{dt} = \sum_j \frac{\partial^2 H}{\partial q_j \partial p_i}\xi_j + \sum_j \frac{\partial^2 H}{\partial p_j \partial p_i}\eta_j, \tag{4}$$

$$(i = 1,...,n)$$

$$\frac{d\eta_i}{dt} = -\sum_j \frac{\partial^2 H}{\partial q_j \partial q_i}\xi_j - \sum_j \frac{\partial^2 H}{\partial p_j \partial q_i}\eta_j, \tag{5}$$

where coefficients are continuous and bounded real functions of *t*. These equations are of essential importance in studies for the stability of motion of the conservative mechanical systems. Let us show this.

Poincaré has found [9] that, if $\xi_s$, $\eta_s$ and $\xi'_s, \eta'_s$ are any two particular solutions of the variational equations (4)-(5), the following invariant takes place

$$\sum_s (\xi_s \eta'_s - \eta_s \xi'_s) = C, \tag{6}$$

where *C* is constant. The proving is elementary and consists in differentiation with respect to *t*.

It is not difficult to show that for each $\xi_s$, $\eta_s$ always exists, at least, one solution $\xi'_s, \eta'_s$ when the constant *C* is nonzero in the Poincaré invariant. Indeed, for the non-trivial solution $\xi_s$, $\eta_s$ some of the initial values $\xi_{s0}$, $\eta_{s0}$ at the time $t_0$ will be different from zero. Then it is always possible to determine the second partial solution by the initial values $\xi'_{s0}, \eta'_{s0}$ so that constant, which is of interest to us, will be nonzero.

Let the constant *C* is nonzero for two solutions $\xi_s$, $\eta_s$ and $\xi'_s, \eta'_s$ of variational equations , and $\lambda$ and $\lambda'$ are the characteristic functions corresponding to these solutions. If to apply the Lyapunov theory of the characteristic values [3] to this invariant, it is possible, on the one hand, to conclude that a characteristic value in the left-hand-side of the invariant (6) corresponding to nonzero constant is zero and, on the other hand, to obtain, as consequence, the following inequality:

$$\lambda + \lambda' \leq 0. \tag{7}$$

Let us suppose now that the system of the Poincaré variational equations is correct. Using the Lyapunov theorems on the solution stability of the systems of differential equations (in the first approximation) [3], it is easy to show that the unperturbed motion of the considered Hamiltonian system is stable when all characteristic values of independent solutions in (6) are equal to zero:

$$\lambda = \lambda' = 0. \tag{8}$$

Thus, Eq. (8) is the condition of stability for the motion of the Hamiltonian system (4)-(5) with respect to the variables $q_i$ and $p_i$ under perturbation of the initial values of variables only, i.e., the constants $\alpha_i$ and $\beta_i$. However, determination of the characteristic values as the function of $\alpha_i$ and $\beta_i$ is a very difficult problem and therefore it is unpractical. The problem becomes simpler in view of the fact that, if the unperturbed motion of our Hamiltonian system under the condition (8) is stable under any perturbations of the initial values, it must be stable under perturbation of the constants $\beta_i$ only. In other words, a problem is reduced to determination of the so-called conditional stability.

Due to this assumption of the nature of initial perturbations we get from the solution of the Hamilton-Jacobi equation (2) (recall $p_i=\partial S/\partial q_i$) the following relations with an accuracy up to the terms of the second order of smallness:

$$\eta_i = \sum_j \frac{\partial^2 S}{\partial q_i \partial q_j} \xi_i, \tag{9}$$

which allows us, taking into consideration the relation

$$H = \frac{1}{2}\sum g_{ij} p_i p_j + U, \tag{10}$$

to rewrite the first group of Eqs. (4) in the form

$$\frac{d\xi_i}{dt} = \frac{1}{2}\sum_{js} \xi_s \frac{\partial}{\partial q_s}\left(g_{ij}\frac{\partial S}{\partial q_j}\right). \tag{11}$$

where coefficients $g_{ij}$ depend on coordinates only. Here the variables $q_j$ and constants $\alpha_j$ on the right-hand side must be substituted for their values corresponding to unperturbed motion.

Now for a stable unpertubed motion (11) be reducible by a nonsingular linear transformation

$$x_i = \sum \gamma_{ij} \xi_j \tag{12}$$

with const determinant $\Gamma=\|\gamma_{ij}\|$. If $\xi_{ir}$ ($r=1,\ldots,n$) are normal system of independent solutions of (11) then (12)

$$x_{ir} = \sum_j \gamma_{ij} \xi_{ir} \tag{13}$$

will be the solution for reduced system. For a stable unpertubed motion all the characteristic values of the solution $x_{ir}(i=1,\ldots,n)$ are zero and consequently taking into account (13) and (11) it is possible to write

$$\|x_{sr}\| = C^* = \|\gamma_{sj}\|\|\xi_{jr}\| = \Gamma C \exp\left[\frac{1}{2}\int \sum \frac{\partial}{\partial q_i}\left(g_{ij}\frac{\partial S}{\partial q_j}\right)dt\right]. \tag{14}$$

As the Lyapunov characteristic value X [$f$] of function $f$ by definition [3] is

$$X[f] = -\overline{\lim_{t\to\infty}}[\log(|f(t)|/t)], \tag{15}$$

then from Eq. (14) for a stable pertubed motion we have [1]

$$\sum_{i,j}\frac{\partial}{\partial q_i}\left(g_{ij}\frac{\partial S}{\partial q_j}\right) = 0, \tag{16}$$

which expresses the vanishing of the sum of characteristic values of the system (11). A simple but elegant proof of condition (16) can be found in Ref. [10].

Let us now consider a more complicated problem. Let a really moving material system undergo the action of both the forces with force function $U$, theoretically considered above, and the unknown perturbation (dissipative) forces which are assumed to be potential and admit the force function $Q$. Then the real motion of the material system occurs in the field of forces with general force function $U^* = U + Q$. In this case, the real motion of the system does not coincide obviously at all with the theoretical one (without perturbation).

If we conserve the statement of the problem on the stability of real nonperturbed motions in the theoretical field of forces with the function $U$ at a perturbation of only initial data as above, then the necessary requirement of stability in the first approximation, e.g., in form (16), will not be efficient in the general case, because the new function $S$ is unknown (as well as $Q$). However, it turns out that we can determine such conditions of stability, which are emplicity independent of the form of the unknown action functions $S$ and the potential $Q$. Thus, we are based on the requirement of stability of form (16), by assuming the conditions of its existence (correctness, etc.) for real motions to be satisfied. In relation (16), we now replace the function $S$ by a new function $\psi$ defined by the equality

$$\psi = A\exp(ikS), \tag{17}$$

where $k$ is a constant; and $A$ is a real function of the generalized coordinates $q$ and the time $t$.

The introduction of a real wave function, like the de Broglie "pilot-wave" [11], is extremely necessary from the physical viewpoint because of the following nontrivial reason. Since the dynamics of a physical system must undoubtedly conserve the Hamilton form of the equations of motion, the main "task" of such a real wave consists in the exact compensation of the action of dissipative forces, which are generated by the perturbation energy $Q$. We will show below that, in this case, such a procedure makes it possible not only to conserve the Hamilton form of the dynamics of a physical system, but allows one to determine the character of an analytic dependence of the energy $Q$ of disturbing forces on the wave function amplitude $\psi$.

Then relation (17) yields

$$\frac{\partial S}{\partial q_j} = \frac{1}{ik}\left(\frac{1}{\psi}\frac{\partial \psi}{\partial q_j} - \frac{1}{A}\frac{\partial A}{\partial q_j}\right), \tag{18}$$

and, hence, relation (16) looks like

$$\sum_{i,j}\frac{\partial}{\partial q_i}\left[g_{ij}\left(\frac{1}{\psi}\frac{\partial \psi}{\partial q_j} - \frac{1}{A}\frac{\partial A}{\partial q_j}\right)\right] = 0. \tag{19}$$

On the other hand, we can write the Hamilton-Jacobi equations for a perturbed motion in the general case where the Hamiltonian $H$ depends explicitly on the time,

$$\frac{1}{2k^2}\sum_{i,j}g_{ij}\left(\frac{1}{\psi}\frac{\partial \psi}{\partial q_i} - \frac{1}{A}\frac{\partial A}{\partial q_i}\right)\left(\frac{1}{\psi}\frac{\partial \psi}{\partial q_j} - \frac{1}{A}\frac{\partial A}{\partial q_j}\right) = \frac{\partial S}{\partial t} + U + Q, \tag{20}$$

where $\partial S/\partial t$ can be determined with the help of (17). Adding relations (19) and (20), we obtain the necessary condition of stability (in the first approximation) in the form

$$\frac{1}{2k^2\psi}\sum_{i,j}\frac{\partial}{\partial q_i}\left(g_{ij}\frac{\partial \psi}{\partial q_j}\right) - \frac{1}{2k^2A}\sum_{i,j}\frac{\partial}{\partial q_i}\left(g_{ij}\frac{\partial A}{\partial q_j}\right) -$$

$$-\frac{1}{k^2A}\sum_{i,j}g_{ij}\frac{\partial A}{\partial q_j}\left(\frac{1}{\psi}\frac{\partial \psi}{\partial q_i} - \frac{1}{A}\frac{\partial A}{\partial q_i}\right) - \frac{1}{ikA\psi}\left[A\frac{\partial \psi}{\partial t} - \psi\frac{\partial A}{\partial t}\right] - U - Q = 0. \tag{21}$$

In this place, we need to use a procedure for the compensation of the action of dissipative forces, which are generated by the perturbation energy $Q$ in order to conserve the Hamilton form of the dynamics of a physical system (21). It is obvious that equality (21) will not contain $Q$, if the amplitude $A$ is determined from the equation

$$\frac{1}{2k^2 A}\sum_{i,j}\frac{\partial}{\partial q_i}\left(g_{ij}\frac{\partial A}{\partial q_j}\right)+\frac{i}{kA}\sum_{i,j}g_{ij}\frac{\partial A}{\partial q_j}\frac{\partial S}{\partial q_i}-\frac{1}{ikA}\frac{\partial A}{\partial t}+Q=0 \qquad (22)$$

which decays into two equations

$$Q=-\frac{1}{2k^2 A}\sum_{i,j}\frac{\partial}{\partial q_i}\left(g_{ij}\frac{\partial A}{\partial q_j}\right), \qquad (23)$$

$$\frac{\partial A}{\partial t}=-\sum_{i,j}g_{ij}\frac{\partial A}{\partial q_j}\frac{\partial S}{\partial q_i} \qquad (24)$$

after the separation of the real and imaginary parts. Here $Q$ is the dissipation energy.

Thus, if the properties of disturbing forces satisfy conditions (23) and (24), then the necessary condition of stability (21) has the form of a differential equation of the "Schrödinger" type:

$$\frac{i}{k}\frac{\partial \psi}{\partial t}=-\frac{1}{2k^2}\sum_{i,j}\frac{\partial}{\partial q_i}\left(g_{ij}\frac{\partial \psi}{\partial q_j}\right)+U\psi, \qquad (25)$$

where $q_1(t),\ldots, q_N(t)$ give the position of the physical system, whose possible trajectories in the $N$-dimensional configuration space $q=(q_1,\ldots,q_N)$ are a solution of the system of the so-called quidance equations [12]

$$\frac{dq_j}{dt}=\frac{1}{m}\nabla_j S, \qquad (26)$$

where $S$ is the phase of the wave function (17).

In other words, we obtained the following result: Eq. (16), which corresponds to the Chetaev stability condition, is transformed into an equation of the "Schrödinger" type (25) with the use of transformation (17). It is obvious that, in the class of equations of type (25), the single-valued, finite, and continuous solutions for the function $\psi$ in the stationary case are admissible only for the eigenvalues of the total energy $E$. Hence, the given stability of real motions takes place only for these values of the total energy $E$. It is worth noting that E. Schrödinger was the first who paid attention (mathematicians knew this for a long time [13]) to such a class of differential equations, in which the fulfilment of such natural conditions as the integrability of the squared modulus of a solution and its finiteness at singular points of the equation [13] is sufficient for the spectrum discreteness (as distinct from, for example, boundary-value problems with boundary conditions).

We now present a short analysis of the obtained results. In our opinion, the plan of Chetaev was as follows. He knew that, according to one of the theorems of stability theory [1], only two types of forces– dissipative and gyroscopic – do not break the stability (if it is present) of a nondisturbed motion of holonomic mechanical systems. Therefore, by introducing a dissipative perturbation $Q$ into the Hamilton--Jacobi equations (20) and by taking simultaneously the stability condition (16) into account, he reasonably expected that, under condition of the conservation of the stability of a mechanical system, it is possible to get a real functional dependence of the dissipation energy $Q$ on characteristics of the wave function $\psi$ (17). Indeed, having obtained the condition for the stability of trajectories of a dynamical system in the form of an equation of the "Schrödinger" type (25), he established not only the physical sense of the perturbation energy $Q$, but he showed also that it is a function of the amplitude of the wave function $\psi$ (17) and takes form (23). It is a very important result. The subsequent content of the article is, as will be clear in what follows, a direct consequence of this result.

It is easy to demonstrate that namely the conclusion about the dissipative nature together with the simultaneous determination of a functional dependence of the energy of disturbing forces on the form (but not on the magnitude (see Eq.(23) of the amplitude of the wave function $\psi$ allow one to generalize an equation of the "Schrödinger" type (25) to the case where the condition of stability (16) is not fulfilled, i.e., $L \neq 0$. Let us show this.

Obvious analysis of Eqs. (16), (19)-(21) shows that the expression

$$\varepsilon = \frac{1}{2k} \sum_{i,j} \frac{\partial}{\partial q_i} \left( g_{ij} \frac{\partial S}{\partial q_j} \right) \qquad (27)$$

with allowance for dimensions is, in general case, the variation of particle kinetic energy predetermined accordingly by variations of its momentum.

Now, if to add and simultaneously to subtract the complex expression ($i\varepsilon$) (see Eq. (27)) and to substitute also Eq. (18) in the left-hand side of the Hamilton-Jacobi equation (7), we get the generalized equation corresponding to the extended Eq. (21):

$$\frac{1}{2k^2\psi}\sum_{i,j}\frac{\partial}{\partial q_i}\left(g_{ij}\frac{\partial\psi}{\partial q_j}\right)-\frac{1}{2k^2A}\sum_{i,j}\frac{\partial}{\partial q_i}\left(g_{ij}\frac{\partial A}{\partial q_j}\right)-\frac{i}{2k}\sum_{i,j}\frac{\partial}{\partial q_i}\left(g_{ij}\frac{\partial S}{\partial q_j}\right)-$$

$$-\frac{1}{k^2A}\sum_{i,j}g_{ij}\frac{\partial A}{\partial q_j}\left(\frac{1}{\psi}\frac{\partial\psi}{\partial q_i}-\frac{1}{A}\frac{\partial A}{\partial q_i}\right)-\frac{1}{ikA\psi}\left[A\frac{\partial\psi}{\partial t}-\psi\frac{\partial A}{\partial t}\right]-U-Q=0. \tag{28}$$

Repeating the ideology of derivation of Eq. (22) it is easy from Eq .(28) to get an equation of the "Schrödinger" type, which is formally identical to Eq. (25), but already under a more general condition imposed on the disturbing energy and the wave function:

$$Q=-\frac{1}{2k^2A}\sum_{i,j}\frac{\partial}{\partial q_i}\left(g_{ij}\frac{\partial A}{\partial q_j}\right), \tag{29}$$

$$\frac{\partial A}{\partial t}=-\frac{A}{2}\sum_{i,j}\frac{\partial}{\partial q_i}\left(g_{ij}\frac{\partial S}{\partial q_j}\right)-\sum_{i,j}g_{ij}\frac{\partial A}{\partial q_j}\frac{\partial S}{\partial q_i}. \tag{30}$$

Physical sense of Eq.(30) for the wave function amplitude consists in the fact that with regard for the formulas for a classical velocity $\bar{v}=\nabla S/m$ and the probability density $P(q, t)=[A(q, t)]^2$ (the substantiation of this formula will be given below) in the configuration space it can be easily transformed into the equation of continuity, which represents the invariability of the total number of "particles" or, in other words, the probability conservation law.

We now return to our problem of quantization on the simplest example. Let us consider a material point with mass $m$ in the field of conservative forces with the force function $U$, which depends, in the general case, on the time. The problem on the stability of motions of such a point will be posed in the Cartesian coordinate system $x_1$, $x_2$, and $x_3$. Denoting the momenta along the axes by $p_1$, $p_2$, and $p_3$, respectively, we obtain that the kinetic energy

$$T=\frac{1}{2m}(p_1^2+p_2^2+p_3^2). \tag{31}$$

In this case, conditions (29) and (30) for the structure of disturbing and compensating forces admit the relations

$$Q=-\frac{\hbar^2}{2m}\frac{\Delta A}{A}, \quad k=\frac{1}{\hbar}, \tag{32}$$

$$\frac{\partial A}{\partial t} = -\frac{A}{2m}\sum\frac{\partial^2 S}{\partial x_i^2} - \sum\frac{\partial A}{\partial x_i}\frac{p_i}{m}, \quad (33)$$

and the differential equation (25) defining stable motions takes the form

$$i\hbar\frac{\partial \psi}{\partial t} = -\frac{\hbar^2}{2m}\Delta\psi + U\psi. \quad (34)$$

That is, it coincides with the Schrödinger equation well known in quantum mechanics [14]. In our case, this equation restricts the choice of the integration constants (the total energy $E$ in the stationary case) in the full Hamilton-Jacobi integral. In what follows, we call Eq. (34) the Schrödinger-Chetaev equation, by emphasizing the specific feature of its origin.

It is of interest to consider the case related to the inverse substitution of the wave function (17) in the Schrödinger equation (34) that generates an equivalent system of equations known as the Bohm-Madelung system of equations [15-17]:

$$\frac{\partial A}{\partial t} = -\frac{1}{2m}[A\Delta S + 2\nabla A \cdot \nabla S], \quad (35)$$

$$\frac{\partial S}{\partial t} = -\left[\frac{(\nabla S)^2}{2m} + U - \frac{\hbar^2}{2m}\frac{\Delta A}{A}\right]. \quad (36)$$

It is very important that the last term in Eq. (36), which is the "quantum" potential of the so-called Bohm $\psi$-field [15-18] in the interpretation of Ref. [15], coincides exactly with the dissipation energy $Q$ in (32). At the same time, Eq. (35) is identical to the condition for $\partial A/\partial t$ in (30) and (33).

If we make substitution of the type

$$P(q,t) = \psi\psi^* = [A(q,t)]^2, \quad (37)$$

then Eqs. (35) and (36) can be rewritten in the form

$$\frac{\partial P}{\partial t} = -\frac{1}{m}\nabla(P \cdot \nabla S), \quad (38)$$

$$\frac{\partial S}{\partial t} + \frac{(\nabla S)^2}{2m} + U - \frac{\hbar^2}{4m}\left[\frac{\Delta P}{P} - \frac{1}{2}\frac{(\nabla P)^2}{P^2}\right] = 0. \quad (39)$$

Here, Eq. (38) has a clear physical sense: $P(q, t)$ is the probability density to find a particle in a certain place of the space, and $\nabla S/m$ is, according to (26), the classical velocity of this particle. In other words, Eq. (38) is nothing but the equation of continuity which indicates how the probability density $P(q, t)$ moves according to the laws of classical mechanics with a classical velocity $\vec{v} = \nabla S/m$ at every point.

On the other hand, we can show that $P(q, t)$ is also the probability density function for the number of particle trajectories, that is substantiated in the following way. We assume that the influence of disturbing forces generated by the potential $Q$ on a wave packet at an arbitrary point of the configuration space is proportional to the density of particle trajectories ($\psi\psi^*=A^2$) at this point. This implies that the disturbing forces do not practically perturb the packet, if the relation

$$\int Q\psi\psi^* dV \Rightarrow \min, \qquad \int \psi\psi^* dV = 1, \tag{40}$$

is satisfied (here, $dV$ stands for an element of the configuration space volume). This means, in its turn, that the disturbing forces admit the absolute stability on the whole set of motions in the configuration space only if condition (37) is satisfied or, in other words, if the following obvious condition of the equivalent variational problem (for $Q$) fulfills:

$$\delta \int Q\psi\psi^* dV = \delta Q = 0. \tag{41}$$

The variational principle (41) is, in essence, the principle of least action of a perturbation. Below, we call it the principle of least action of perturbations by Chetaev [1,19].

Using the previous notations, we write the following equality for $Q$:

$$Q = -\frac{\partial S}{\partial t} - U - T = -\frac{\partial S}{\partial t} - U - \frac{1}{2}\sum_{ij} g_{ij} \frac{\partial S}{\partial q_i} \frac{\partial S}{\partial q_j}. \tag{42}$$

On the other hand, if (4) holds true, it is easy to show that

$$\frac{1}{2}\sum_{ij} g_{ij} \frac{\partial S}{\partial q_i} \frac{\partial S}{\partial q_j} = -\frac{1}{2k^2\psi^2}\sum_{ij} g_{ij} \frac{\partial \psi}{\partial q_i} \frac{\partial \psi}{\partial q_j} + \frac{1}{2k^2 A^2}\sum_{ij} g_{ij} \frac{\partial A}{\partial q_i} \frac{\partial A}{\partial q_j} +$$

$$+ ik\frac{1}{2k^2 A^2}\sum_{ij} g_{ij} \frac{\partial A}{\partial q_i} \frac{\partial S}{\partial q_j}. \tag{43}$$

Then it is necessary to perform the following subsequent substitutions. First, we substitute relation (43) in (42) and then introduce the result in the equation corresponding to the variational principle (41).

As a result of the indicated procedure of substitutions, we obtain the relation which is exactly equal to Eq. (28), i.e., to the extended equation (with regard for (16) and (27)] of type (21). Hence, the structural expressions and the necessary condition of stability which follow from it coincide with (29), (30), and (25), respectively. This means that, on the basis of the Chetaev variational principle (28), we get the independent confirmation of the fact that the physical nature of $P(q, t)$ reflects really not only the traditional notion of the probability density for a particle to be at a certain place of the space according to the Bohm-Madelung equation of continuity (25) but plays also the role adequate to that of the probability density of the number of particle trajectories.

Such semantic content of the probability density function $P(q, t)$ and simultaneously the exact coincidence of the "quantum" potential of the Bohm $\psi$-field [15-18] in Eq. (36) and the force function of perturbations $Q$ in (32) lead immediately to surprising, but fundamental conclusions:

- in the light of the Chetaev theorem on stable trajectories in dynamics, the reality of the Bohm $\psi$-field is the obvious firmly established fact which leads, in its turn, to a paradoxical, at first glance, conclusion that classical dynamics and quantum mechanics are, in fact, two mutually complementing procedures of the single Hamilton theory. In other words, classical dynamics and the conditions of the quantization (stability) are, contrary to the well-known correspondence principle, two mutually complementary procedures of a description of the stable motion of a classical physical system in the field of potential forces. Moreover, in the framework of this theory, the quantum equation (36) is an ordinary Hamilton-Jacobi equation and differs from an analogous classical equation at $\hbar \to 0$ ($Q \to 0$ [16]) by that its solution is a priori stable. From this viewpoint, it is natural that namely this difference is the reason for such known phenomenon as the quantum chaos, which characterizes the specific features of quantum-mechanical systems chaotic in the classical limit [20];

− it is obvious that, in the light of the Chetaev theorem, the sense of the Heisenberg uncertainty relations is cardinally changed. In this case, the main reason for the statistical scattering, which is characterized by dispersions of coordinates and momenta, is small dissipative forces. These forces are generated by a perturbation potential or, what is the same, by quantum potential $Q$. In this case, it is easy to show that the dispersions of coordinates and momenta are determined by the averaged quantum potential $\langle Q \rangle$. This is visually demonstrated in the framework of a mathematical representation of the uncertainty relation in the one-dimensional case in the following form (see Eq. (6.7.23) in [16])

$$\langle(\Delta x)^2\rangle\langle(\Delta p_x)^2\rangle = \langle(\Delta x)^2\rangle\langle Q\rangle 2m \geq \hbar^2/4, \tag{44}$$

where $\langle(\Delta x)^2\rangle$ and $\langle(\Delta p_x)^2\rangle$ are the dispersions of coordinates and momenta;

− on the basis of the principle of least action of perturbations (41), it is shown that the function *P(q, t)* is equivalent semantically and syntactically to the probability density function of the number of trajectories of a particle. This result means that the above-given proof (see relations (40)-(43)) for the function *P(q, t)* jointly with the new (Chetaev's) interpretation of the Heisenberg uncertainty relation are the direct unique substantiation of the very important fact that the Bohm quantum mechanics supplemented by the Chetaev generalized theorem does not basically contain the hidden parameters in the form of the velocity and the coordinate of a particle. In other words, these parameters characterizing a trajectory of the particle not only exist, but they are completely determined by a wave solution of the Schrödinger-Chetaev equation (34) under condition (26) or, to be more exact, by the probability density function of the number of particle trajectories;

− it is obvious that the Bohm-Chetaev interpretation of quantum mechanics is a constructive generalization of the essence of the nonlocal Bohm quantum mechanics [15-17], within which it was shown that the Schrödinger equation together with the Born probabilistic postulate (which is mathematically equivalent to formula (37)) include the adequate description of the whole experimental-theoretical contents of modern quantum mechanics. The last means that the Bohm-Chetaev quantum theory with trajectories is physically equivalent to the traditional quantum mechanics.

It is worth noting that all most known alternative theories of quantum mechanics are nonlocal theories. In the first turn, this concerns directly the Bohm mechanics [15-17], the Girardi-Rimini-Weber model of a united dynamics of micro- and macroscopic systems [21] and its relativistic version [22], and a number of the well-known quantum models which were developed exclusively on the basis of classical notions: the Nelson stochastic quantum mechanics [23], the Nottale fractal quantum mechanics [24], the Santamato geometric quantum mechanics [25], the Grössing "thermodynamical" quantum mechanics [26] and the Wetterich "statistical" quantum mechanics [27]. It is worth noting that, due to their nonlocality, these theories, including the Bohm-Chetaev quantum theory with trajectories, do not contradict the so-called Bell inequalities [28,29], whose violation was unambiguously corroborated by numerous high-precision experiments.

At the same time, it is necessary to remember that the essential difference of the Bohm-Chetaev quantum mechanics from the mentioned alternative theories is the unconditional identity of the probability density function of the number of trajectories of a particle obtained on the basis of the principle of least action of perturbations (41) and the probability density function for a particle to be in a certain place of the configuration space obtained on the basis of the Bohm-Madelung equation of continuity (38). This exclusively important fact emphasizes naturally the physical identity of the probabilistic and trajectory interpretations of quantum mechanics.

We now consider the fundamental differences of the Bohm quantum mechanics supplemented by the generalized Chetaev theorem from the Bohm quantum mechanics itself and from the traditional, i.e. probabilistic, quantum mechanics.

The obvious analysis of the Schrödinger-Chetaev equation (34) as the condition of stability of trajectories of a classical physical system (a particle) in the field of potental forces puts naturally the extremely profound fundamental question about the physical nature of really existent, as was shown above, small disturbing forces or "small dissipative forces with the full dissipation" by Chetaev [2] on the first plan. It is obvious that we deal with perturbation waves of the de Broglie type, whose action is described by the Bohm $\psi$-field. Such a conclusion is determined, in the first turn, by that the "embryonic theory of a union of waves and particles" by de Broglie [10, 30] was constructed just on the basis of the identity of the Hamilton principle of least action and the Fermat principle, which reflects very exactly and clearly the physical essence of the Chetaev theorem on stable trajectories in dynamics (see Eq. (41)). Moreover, in the framework of the Bohm mechanics supplemented by the generalized Chetaev theorem, it is easy to conclude that the reality or, to be more exact, the observability of de Broglie wave is ensured, first, by the reality of the Bohm $\psi$–field which has the sense of the dissipation energy $Q$ in (32) and (36) and, second, is thedirect consequence of the absence of hidden parameters. In other words, the dissipative reality of the quantum potential and the absence of hidden parameters in the Bohm-Chetaev quantum mechanics are, respectively, the necessary and sufficient conditions for the reality of de Broglie waves.

Thus, we may conclude that the Bohm-Chetaev quantum mechanics as well as the probabilistic quantum mechanics, is, on the one hand, a theory without hidden parameters and, on the other hand, a nonlocal theory. Hence, due to the indicated reasons, the main essential difference of these two theories is the fact of the physical (non)observability of de Broglie waves. If a wave is not observable physically, then the probabilistic interpretation of quantum mechanics is valid. On the contrary, if the experiment will prove the real existence of de Broglie waves, then, in this case, the Chetaev interpretation of quantum mechanics is proper.

It is worth noting that the question about the reality of de Broglie waves has a long history and is not exotic or metaphysical at present. In this aspect, it becomes clear with regard for the previous experience that the execution of new fundamental experiments should be based on the principle of direct registration of the real wave function with the help of a supersensitive registration of the interference of (electromagnetic? [10,23,30,31]) waves of perturbations (trajectories) which accompany the diffraction of electrons (or neutrons) from a low-intensity source, like that in the experiment by Tonomura *et al*. [32]. This question becomes especially actual in connection with new data obtained by Catillon *et al.* [33] in the experiments on the channeling of electrons in a thin Si crystal. They obtained the results testifying to the anomalous scattering which indicate the possibility of the stable motion of electrons of the *zitterbewegung* type [16,31]. It is obvious that this result, if it will be reliably confirmed, is simultaneously the direct indication of the material reality of de Broglie waves [30].

## ACKNOWLEDGMENT

The authors acknowledge the financial support of the DFG through SFB 701 "Spectral structures and topological methods in mathematics", Bielefeld University (Germany), German-Ukrainian Project 436 UKR 113/94, and IGK Bielefeld (Germany).